*Sequence analysis*

# SOAPdenovo-Trans: *De novo* transcriptome assembly with short RNA-Seq reads

Yinlong Xie[1,2,3,†], Gengxiong Wu[1,†], Jingbo Tang[1,4,†], Ruibang Luo[1,2,6,†], Jordan Patterson[5], Shanlin Liu[1], Weihua Huang[1], Guangzhu He[1], Shengchang Gu[1], Shengkang Li[1], Xin Zhou[1], Tak-Wah Lam[2], Yingrui Li[6], Xun Xu[1], Gane Ka-Shu Wong[1,5,7,*], Jun Wang[1,8,9,10,*]

[1]BGI-Shenzhen, Shenzhen, China.
[2]HKU-BGI Bioinformatics Algorithms and Core Technology Research Laboratory & Department of Computer Science, University of Hong Kong, Pokfulam, Hong Kong.
[3]School of Bioscience and Bioengineering, South China University of Technology, Guangzhou, China.
[4]Institute of Biomedical Engineering, XiangYa School of Medicine, Central South University, Changsha, China.
[5]Department of Medicine, University of Alberta, Edmonton AB, Canada T6G 2E1.
[6]BGI-tech, BGI-Shenzhen, Shenzhen, China.
[7]Department of Biological Sciences, University of Alberta, Edmonton AB, Canada T6G 2E9.
[8]The Novo Nordisk Foundation Center for Basic Metabolic Research, University of Copenhagen, Copenhagen, Denmark.
[9]Department of Biology, University of Copenhagen, Copenhagen, Denmark.
[10]King Abdulaziz University, Jeddah, Saudi Arabia.
[†]Joint first authors.

**ABSTRACT**
**Motivation:** Transcriptome sequencing has long been the favored method for quickly and inexpensively obtaining the sequences for a large number of genes from an organism with no reference genome. With the rapidly increasing throughputs and decreasing costs of next generation sequencing, RNA-Seq has gained in popularity; but given the typically short reads (e.g. 2 × 90 bp paired ends) of this technology, *de novo* assembly to recover complete or full-length transcript sequences remains an algorithmic challenge.
**Results:** We present SOAPdenovo-Trans, a *de novo* transcriptome assembler designed specifically for RNA-Seq. Its performance was evaluated on transcriptome datasets from rice and mouse. Using the known transcripts from these well-annotated genomes (sequenced a decade ago) as our benchmark, we assessed how SOAPdenovo-Trans and two other popular software handle the practical issues of alternative splicing and variable expression levels. Our conclusion is that SOAPdenovo-Trans provides higher contiguity, lower redundancy, and faster execution.
**Availability and Implementation:** Source code and user manual are at http://sourceforge.net/projects/soapdenovotrans/
**Contact:** xieyl@genomics.cn or bgi-soap@googlegroups.com

**Supplementary information:** Supplementary information is available at ftp://public.genomics.org.cn/BGI/SOAPdenovo-Trans

## 1 INTRODUCTION

With the rapidly increasing throughputs and decreasing costs of next generation sequencing, RNA-Seq has become an efficient way to get information on gene expression levels, as well as on the transcript sequences themselves. The latter applications include discriminating expression levels of allelic variants and detecting gene fusions (Maher, et al., 2009). Assemblers like Cufflinks (Trapnell, et al., 2010), Scripture (Guttman, et al., 2010), and ERANGE (Mortazavi, et al., 2008) recover transcript sequences by aligning the reads to a reference genome. However, reference genomes are not always available, especially if the genome is unusually large and/or polyploid, as is often the case for plant genomes. *De novo* assembly is then a favorable choice. The challenge is not only to recover full-length transcripts, but also to identify all alternative splice forms in the presence of variable gene expression levels. Most contemporary *de novo* transcriptome assemblers, like Trans-ABySS (Robertson, et al., 2010), Multiple-k (Surget-Groba and Montoya-Burgos, 2010), Rnnotator (Martin, et al., 2010), Oases (Schulz, et al., 2012) and Trinity (Grabherr, et al., 2011) utilize

[*]To whom correspondence should be addressed.

the *de Bruijn* graph (DBG) schema for computational and memory efficiency.

Historically, the first assemblers for next generation sequences, like Velvet (Zerbino and Birney, 2008), ABySS(Simpson, et al., 2009) and SOAPdenovo (Li, et al., 2010), were developed for whole genome assembly. They anticipated having to recover tens of chromosomes. In contrast, transcriptome assemblers must recover tens of thousands of RNA sequences. Transcript sequences are typically only a (k)ilobase in length, as compared to chromosomes that can be up to hundreds of (M)egabases in length. Critically, gene expression levels can vary by many orders of magnitude, so that for any nonzero sequencing error rate the most highly expressed genes will always harbor many discrepant bases, making it impossible to define an absolute threshold for the number of sequencing errors allowed per assembly. Further complicating the situation, alternative splice forms transcribed from the same locus intermix into a single complicated *de Bruijn* sub-graph. Oases enumerated all possible transcripts with relatively simple assembly sub-graphs and then used a robust heuristic algorithm to traverse these graphs. Trinity introduced a new error removal model to account for variations in gene expression levels and then used a dynamic programming procedure to traverse their graphs. However, both algorithms have shortcomings, as will be apparent in the performance comparisons to be presented. Oases produces more redundant transcripts, perhaps because it lacks an effective error removal model (Lu, et al., 2013). Trinity produces fewer full-length transcripts, because it does not use paired-end data for scaffold construction.

We present a *de novo* RNA-Seq assembler, SOAPdenovo-Trans, that incorporates the innovations of these previous algorithms and then improves on them with subtle changes learned from years of practical experience at BGI. It adopts the error removal model from Trinity and the robust heuristic graph traversal method from Oases. In addition, using a strict transitive reduction method, we simplify the scaffolding graphs more accurately. We evaluated all of these assemblers on transcriptome data from rice and mouse, using the known transcripts from these well-annotated genomes (sequenced a decade ago) as our benchmark. The results demonstrate that SOAPdenovo-Trans provides higher contiguity, lower redundancy, and faster execution.

## 2 METHODS

SOAPdenovo-Trans is a *de Bruijn* graph-based assembler for *de novo* transcriptome assembly, derived from the SOAPdenovo2 (Luo, et al., 2012) genome assembler, which had an effective scaffolding module that is just as relevant for transcriptome assembly. However, SOAPdenovo2 was designed for genomes with uniform sequencing depth. The error removal model that they used is not applicable to RNA-Seq data. Moreover, they did not allow for alternative splicing. By adopting and improving on concepts from Trinity and Oases, both of these problems were solved.

An overview of the SOAPdenovo-Trans algorithm is shown in Fig. 1A. It consists of two main steps that we will describe under the headings contig assembly and transcript assembly.

### 2.1 Contig assembly

*De Bruijn* graph construction is done following SOAPdenovo2, but sequencing errors are removed in two ways: globally (also done by the genome version of SOAPdenovo2) and locally (only done by this transcriptome version). In global error removal, low-frequency *k*-mers, edges, arcs (direct linkage between contigs in the *de Bruijn* graph) and tips are removed; bubbles are pinched. The assumption is that most are sequencing errors. However, for the most highly expressed genes, the inevitable multitude of sequencing errors will always exceed any global error removal threshold. Hence, we only apply a weak depth cutoff (by default ≤2) in global error removal in order not to mistakenly remove low expressed genes from the graph. Following the initiative from Trinity, we rely more on local error removal to deal with sequencing errors. The idea is to apply a percentage threshold for filtration (say ≤5% of the total or maximal depth of the adjacent graph element, which can refer to *k*-mers, arcs or edges) instead of a constant threshold. Lastly, we generate contigs in the manner of SOAPdenovo2.

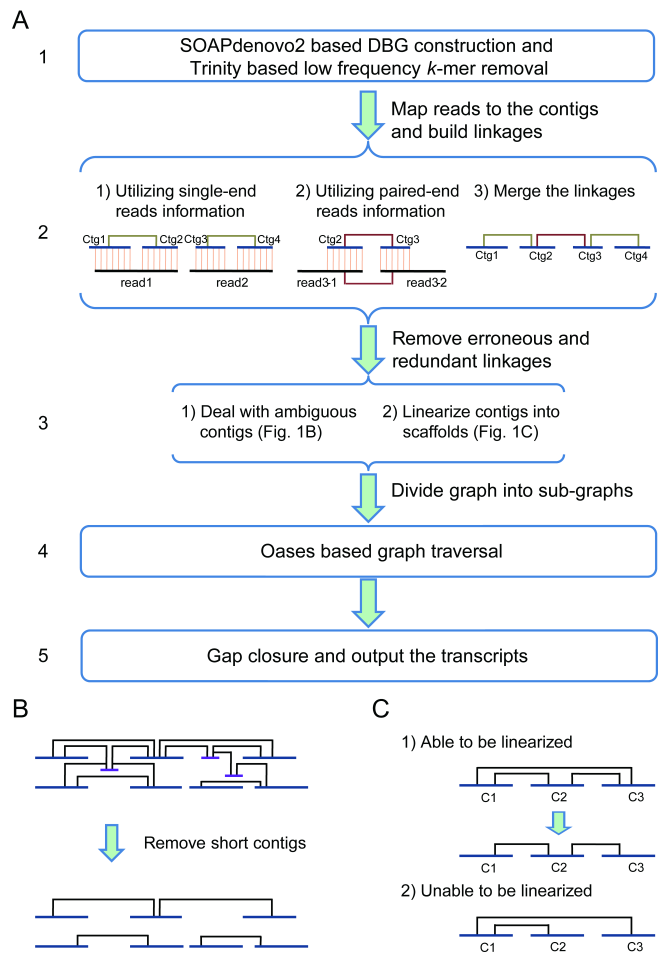

**Fig. 1.** Overview of SOAPdenovo-Trans algorithm. (A1) Contig assembly. *De Bruijn* graphs are constructed from reads, sequencing errors are removed, and contigs are then constructed. (A2-A5) Transcript assembly. Single-end and paired-end reads are mapped to the assembled contigs to construct scaffold graphs. Transcripts are generated by traversing through reliable paths for each graph. (B) Dealing with ambiguous contigs. (C) Linearizing contigs into scaffolds.

### 2.2 Transcript assembly

*2.2.1 Scaffold construction* Reads are mapped back onto the contigs to build linkages. This is similar to SOAPdenovo2, with the difference being that SOAPdenovo-Trans uses both single-end and paired-end reads information, while SOAPdenovo2 used only paired-ends. This is important because transcripts are much shorter than chromosomes, so we cannot afford to ignore the information that may only be found in single-end reads. The number of reads is then used to assign weights to these linkages, and insert sizes from the paired-ends estimate the distances between linkages.

*2.2.2 Graph simplification* Contigs that are identified as being ambiguous, with multiple successive linkages, or as of exceptionally high depth, with say two times the mean depth, were masked for scaffold building in the genome version of SOAPdenovo2. This is inappropriate for transcriptome assembly because of the existence of alternative splicing and variable gene expression levels. Alternative splicing establishes multiple successive linkages from a unique contig, the data representation of which is analogous to ambiguities in whole genome assembly. Variable gene expression levels make it impossible to define a contig as repetitive using a single depth constant. One of the methods by which SOAPdenovo-Trans copes with these problems is it removes short contigs (default ≤100 bp) unconditionally. This removes not only sequencing errors, but also, short ambiguous contigs caused by repeats, which in turn liberates the scaffolding module from having to solve complicated ambiguities, thereby increasing its ability to identify real instances of alternative splicing (Fig. 1B). Removing short contigs unconditionally will obviously create many small gaps, but this is corrected in the final phase of the algorithm, using a gap-filling module.

Linearization of contigs to scaffolds also differs in genome and transcriptome assembly. For genomes, after introducing paired-end reads with multiple tiers of insert sizes, a starting contig may have multiple successive contigs at different distances from the starting contig. These contigs are expected to be linearly integrated into a single scaffold. However, for transcriptomes, conflicts may legitimately arise from multiple alternative splice forms sharing the same starting contig. To simplify the graphs correctly, SOAPdenovo-Trans utilizes a more stringent linearization method (Fig. 1C): For example, there are three contigs named $c_1$, $c_2$, $c_3$. These can be linearized if 1) there exists explicit linkages between "$c_1$ and $c_2$", "$c_2$ and $c_3$", and "$c_1$ and $c_3$"; 2) the distances between $c_1$, $c_2$ and $c_3$ inferred from linkages do not conflict each other.

*2.2.3 Graph traversal* Contigs are clustered into sub-graphs according to the linkages. Each sub-graph comprises a set of transcripts (alternative splice forms) sharing common exons. SOAPdenovo-Trans traverses these sub-graphs with the algorithm introduced by Oases to generate possible transcripts from linear, fork and bubble paths. For the most complex paths, only the top scoring transcripts are kept.

*2.2.4 Gap filling/correction* Recall that many small gaps were introduced by masking contigs ≤100 bp before scaffold construction. To compensate for this, we used the *de Bruijn* based gap filling method from SOAPdenovo2. Paired-end information is used to cluster semi-unmapped reads into the gap regions, and then we locally assemble these reads into a consensus. In instances where multiple consensus sequences are assembled, we break the tie by selecting the one whose length is most consistent with the gap size.

**2.3 Benchmark to genome**

The rice transcriptome dataset is from *Oryza sativa 9311* (panicle at booting stage). Paired-end sequences were generated on the Illumina GA platform (Zhang, et al., 2010) with 200 bp insert sizes and 75 bp read lengths. What we call the (L)arge dataset is 39.9M reads totaling 5.98 Gbp of sequence.

http://www.ncbi.nlm.nih.gov/sra/SRX017631
http://www.ncbi.nlm.nih.gov/sra/SRX017632
http://www.ncbi.nlm.nih.gov/sra/SRX017633
http://www.ncbi.nlm.nih.gov/sra/SRX017630

What we call the (S)mall dataset is 9.8M reads totaling 1.47 Gbp that we down-sampled from the above.

The mouse transcriptome dataset is from *Mus musculus* (dendritic cells). Paired-end sequences were generated on the Illumina GAII platform (Grabherr, et al., 2011) with 300 bp insert sizes and 76 bp read lengths. What we call the (L)arge dataset is 36.1M reads totaling 5.49Gbp (after quality filtering).

http://www.ncbi.nlm.nih.gov/sra/SRX062280

What we call the (S)mall dataset is 12.0M reads totaling 1.83Gbp that we down-sampled from the above.

As Trinity only supports 25-mers, all assemblers were run with *k*-mer=25, to make the comparisons more equitable. SOAPdenovo-Trans (version 0.99) was run with parameters "-i 20 -q 5 -Q 2 -H 200 -e 20 -S 48 –r -F -L 100 -c 2 -t 5". Oases (version 0.1.21) with Velvet (version 1.1.03) was run using minimum-length-of-output-transcripts set to 100. Trinity (version r2011-08-20) was run with minimum-assembled-contig-length-to-report set to 100. The reference genomes and curated annotations were downloaded from the following two websites.

Rice: MSU Rice Genome Annotation Project Release 7 at
ftp://ftp.plantbiology.msu.edu/pub/data/Eukaryotic_Projects/o_sativa/annotation_dbs/pseudomolecules/version_7.0/all.dir/all.cdna

Mouse: Mus_musculus.NCBIM37.64 at
ftp://ftp.ensembl.org/pub/release-64/fasta/mus_musculus/cdna/Mus_musculus.NCBIM37.64.cdna.all.fa.gz

Note that for rice, our transcriptomes came from the *indica* subspecies, where as our genome came from the *japonica* subspecies. We chose the *japonica* genome because more work has been done on the annotations. However, we had to use *indica* transcriptomes because there is little *japonica* data on the Illumina platform that is freely available. Although the subspecies differ by only a fraction of a percent (Yu, et al., 2005), on average, there are local regions of higher variability that prevent some *indica* transcripts from aligning to the *japonica* genome.

All of the transcript-to-genome alignments are done in BLAT (Kent, 2002) using a 95% identity cutoff. We require that 95% of the transcript length be accounted for in one consistent alignment before we deem the transcript to be correctly assembled. When that fails, we search for "chimeric" assemblies that account for 95% of the transcript length with multiple alignments in different orientations, on different chromosomes, or in distal regions of the same chromosome. When a transcript can be aligned to multiple genome loci, we choose the locus with the longest alignment. No effort was made to determine the "best" alignment when different genome loci give the same aligned length, since this was found in less than 1% of the assemblies. When multiple transcripts align to the same genome locus, and we need a single representative, we choose the largest of these (putative) alternative splice forms.

**Table 1.** Computational requirements.

| Method | Rice | | | | Mouse | | | |
|---|---|---|---|---|---|---|---|---|
| | Small dataset | | Large dataset | | Small dataset | | Large dataset | |
| | Peak memory (GB) | Time (hr) | Peak memory (GB) | Time (hr) | Peak memory (GB) | Time (hr) | Peak memory (GB) | Time (hr) |
| SOAPdenovo-Trans | 10.7 | 0.2 | 29.3 | 0.8 | 10.5 | 0.3 | 16.7 | 1.0 |
| Trinity | 10.1 | 17.7 | 37.6 | 35.6 | 10.5 | 16.6 | 26.3 | 47.5 |
| Oases | 9.0 | 0.6 | 53.2 | 3.0 | 8.8 | 0.7 | 35.1 | 2.7 |

All assemblies were processed with 10 threads, on a computer with two Quad-core Intel 2.8GHz CPUs and 70GB of memory, running CentOS 5.

## 3 RESULTS

To assess the performance of SOAPdenovo-Trans, Trinity and Oases, we assembled two sets ("Large" and "Small") of paired-end Illumina data for rice and mouse. We chose a plant and an animal because we know plants are especially challenging (*i.e.* they have larger gene families, more transposable elements (TEs), and some of these TEs are highly expressed too). It was important for our software to deal with these difficulties because it was designed for use in the 1000 plants (1KP) initiative www.onekp.com. As both genomes were sequenced a decade ago, the annotations have been extensively curated, and are appropriate benchmarks by which the assembly software can be judged. Table 1 shows the computational demands in peak memory and time. SOAPdenovo-Trans is clearly competitive on both issues.

Table 2 summarizes the alignment of the assembled transcripts to the annotated genomes. SOAPdenovo-Trans produces the fewest transcripts, by more than a factor of two in the extreme case, even after removing assemblies below 300 bp. The number of annotated genome loci recovered is surprisingly consistent, differing by only a few percent between algorithms. One might therefore conclude that the large difference in transcript numbers is due to alternative splice forms. However, just because multiple assembled transcripts align to the same genome locus, that does not by itself prove that these are genuine alternative splice forms. They can, for example, be two different fragments of the same isoform or redundant copies of the same isoform.

In recognition of the extensively curated annotations for the rice and mouse genomes, all discussions from here on will be restricted to transcripts that align to genome loci with annotated genes. We use the terms series-A and series-B to denote the sets of transcripts that include and exclude putative alternative splice forms, respectively. Before discussing how the assemblers differ, we must first point out how rice and mouse differ, so that these issues do not become a distraction later on.

Given similar amounts of raw input data, *i.e.* S versus L datasets, the rice assemblies produced more genes than mouse, ranging from 42% to 70% based on the series-B gene counts. This increase cannot be explained by the abundance of transposable elements (TEs) in rice, since more than 95% of the expressed rice genes are non-TEs. Because so many more rice genes must be recovered from the same amount of data, the read depths per gene tend to be lower, so the assemblies will never be as good as with mouse. One might also expect that, in the absence of grotesque assembly errors (*i.e.* so bad that they are not even recognized as chimeric), all but a few percent of the assembled transcripts should align to the reference genome. This was not the case for rice, where close to 10% failed to align because of subspecies differences arising from our use of *indica* transcriptomes with *japonica* genome annotations. By aligning to the combined genomes of both subspecies, the problem disappeared. However, to simplify matters for this paper, we only use the alignments to the *japonica* genome.

When comparing assembled to annotated transcripts, we cannot assume that their exons will overlap perfectly, even if both align to the same genome locus. Indeed this is expected to happen if they are from different isoforms. In most cases however, perfect overlap is observed, as we show in Fig. 2. Hence, despite the possibility of alternative splicing forms, the assembled and annotated transcripts usually represent much the same exons. This allows us to simplify the next plot, Fig. 3, which shows the cumulant for the assembled transcript lengths versus the assembled-to-annotation length ratios. It shows the extent to which full-length transcripts are recovered at whatever level of completeness is desired. The ideal is a step function with a rapid increase at ratios near 100%. SOAPdenovo-Trans comes closer to this ideal than either Trinity or Oases. Based on the "shoulder" in the curve, one can argue that SOAPdenovo-Trans with only 1.83 Gbp of mouse data outperform Trinity with 5.49 Gbp of mouse data.

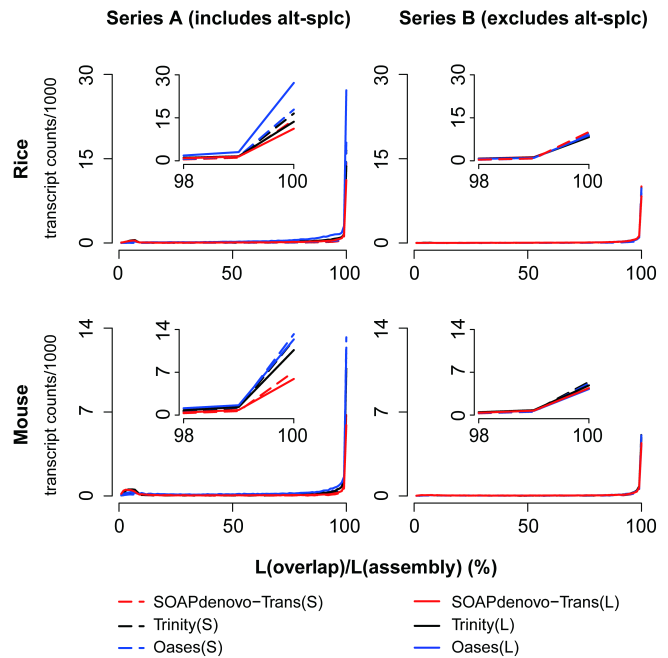

**Fig. 2.** Overlaps between the assembly and annotation. L(overlap) is the length of overlap between the assembled and annotated transcripts, while L(assembly) is the length of the assembled transcript counting only the portion that successfully aligned to the genome. Here we show a distribution in the number of assembled transcripts as a function of the overlap-to-assembly lengths.

**Table 2.** Classification of assembled transcripts. Our analyses generate a successive reduction in the number of assemblies. To begin, we restrict to assemblies larger than 300 bp. BLAT alignments to the reference genomes are done at 95% sequence identity. Assemblies are deemed to be correct when ≥95% of their lengths can be accounted for in one consistent alignment. If not, assemblies are deemed to be "chimeric" when 95% of their lengths can be accounted for in two or more alignments with different orientations, on different chromosomes, or on distal regions of the same chromosome. We then restrict to assemblies that overlap with annotated genes. Because multiple assemblies can align to the same genome locus, we generated two datasets: series-A and series-B, which include and exclude putative alternative splicing forms, respectively. In choosing among isoforms, whether for series B or the genome annotations, we always use the longest available sequence. In the case of the rice transcriptome, about 30.3% of the 55,986 annotated genome loci are known to be TEs, and to show that this is not a confounding issue, we indicate the percentage of the assembled transcripts that are not known to be TEs.

| | Rice | | | | | | Mouse | | | | | |
|---|---|---|---|---|---|---|---|---|---|---|---|---|
| | Small dataset | | | Large dataset | | | Small dataset | | | Large dataset | | |
| | SOAPdenovo-Trans | Trinity | Oases | SOAPdenovo-Trans | Trinity | Oases | SOAPdenovo-Trans | Trinity | Oases | SOAPdenovo-Trans | Trinity | Oases |
| All sizes | 61,425 | 108,946 | 64,445 | 99,398 | 167,187 | 174,738 | 48,224 | 96,400 | 69,090 | 86,961 | 172,425 | 79,046 |
| >300 bp | 25,800 | 35,690 | 36,145 | 38,789 | 54,546 | 92,381 | 16,286 | 29,449 | 34,767 | 25,037 | 43,892 | 49,624 |
| Correct | 23,682 | 29,835 | 30,051 | 34,718 | 44,148 | 75,866 | 15,959 | 27,665 | 32,729 | 24,318 | 40,722 | 46,171 |
| Correct % | 91.80% | 83.60% | 83.10% | 89.50% | 80.90% | 82.10% | 98.00% | 93.90% | 94.10% | 97.10% | 92.80% | 93.00% |
| Chimeric | 526 | 2,045 | 2,158 | 1,020 | 4,124 | 5,102 | 170 | 1,223 | 982 | 439 | 2,378 | 1,836 |
| Chimeric % | 2.0% | 5.7% | 6.0% | 2.6% | 7.6% | 5.5% | 1.0% | 4.2% | 2.8% | 1.8% | 5.4% | 3.7% |
| Series A (includes AS) | 21,557 | 26,870 | 27,675 | 27,965 | 35,394 | 66,158 | 13,032 | 21,631 | 26,592 | 16,789 | 26,766 | 35,130 |
| Series A (nonTE) | 20,614 | 25,601 | 26,444 | 26,698 | 33,363 | 63,249 | - | - | - | - | - | - |
| Series A % (nonTE) | 95.6% | 95.3% | 95.6% | 95.5% | 94.3% | 95.6% | - | - | - | - | - | - |
| Series B (excludes AS) | 14,651 | 14,120 | 13,413 | 17,656 | 16,490 | 16,880 | 9,442 | 9,568 | 9,425 | 10,409 | 10,488 | 10,016 |
| Series B (nonTE) | 14,090 | 13,556 | 12,895 | 16,874 | 15,695 | 16,058 | - | - | - | - | - | - |
| Series B % (nonTE) | 96.2% | 96.0% | 96.1% | 95.6% | 95.2% | 95.1% | - | - | - | - | - | - |

**Table 3.** Evaluations based on number of "full-length" annotations recovered. The alignment criterion is at least 95% sequence identity covering the entire (or ≥95%) annotation and containing at most 5% insertions or deletions.

| | Rice | | | | | | Mouse | | | | | |
|---|---|---|---|---|---|---|---|---|---|---|---|---|
| | Small dataset | | | Large dataset | | | Small dataset | | | Large dataset | | |
| | SOAPdenovo-Trans | Trinity | Oases | SOAPdenovo-Trans | Trinity | Oases | SOAPdenovo-Trans | Trinity | Oases | SOAPdenovo-Trans | Trinity | Oases |
| Coverage = 100% | | | | | | | | | | | | |
| Genes | 386 | 414 | 356 | 1,589 | 1,386 | 883 | 2,897 | 3,015 | 3,064 | 4,303 | 4,265 | 4,175 |
| Isoforms | 405 | 449 | 382 | 1,769 | 1,589 | 980 | 3,505 | 3,807 | 4,051 | 5,572 | 5,786 | 6,300 |
| Coverage ≥ 95% | | | | | | | | | | | | |
| Genes | 1,904 | 1,504 | 1,482 | 5,103 | 3,747 | 2,876 | 6,000 | 5,046 | 5,648 | 7,963 | 6,533 | 7,163 |
| Isoforms | 2,300 | 1,869 | 1,874 | 6,237 | 4,662 | 3,603 | 9,043 | 7,474 | 9,177 | 12,663 | 10,303 | 13,018 |

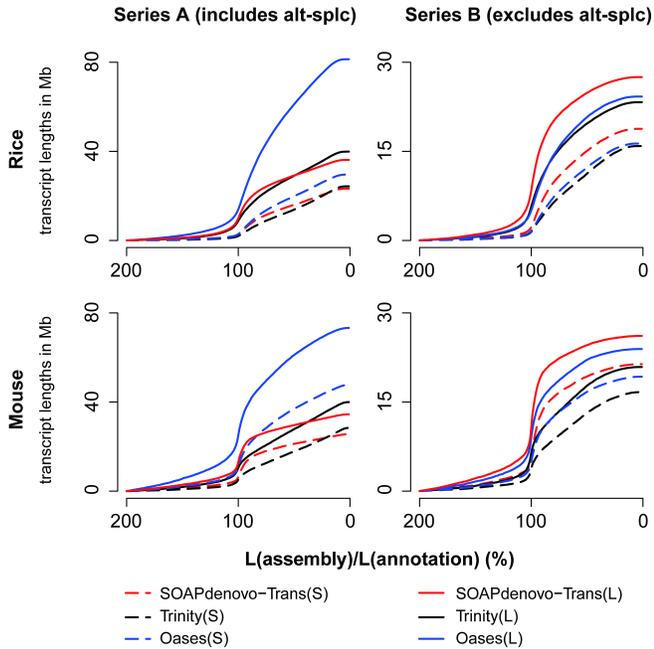

**Fig. 3.** Cumulants of assembled transcript lengths. In contrast to Figure 2 where we showed a distribution, here we plot a cumulant. L(assembly) is the length of the assembled transcript, counting only the portion that aligned to the genome, while L(annotation) is the length of the annotated transcript. Notice that the assembly-to-annotation lengths are plotted in reverse, from large to small. The ideal result is a step function with a sharp rise at 100%; but it starts to rise before one because assembled transcripts contain UTRs not found in many annotated transcripts.

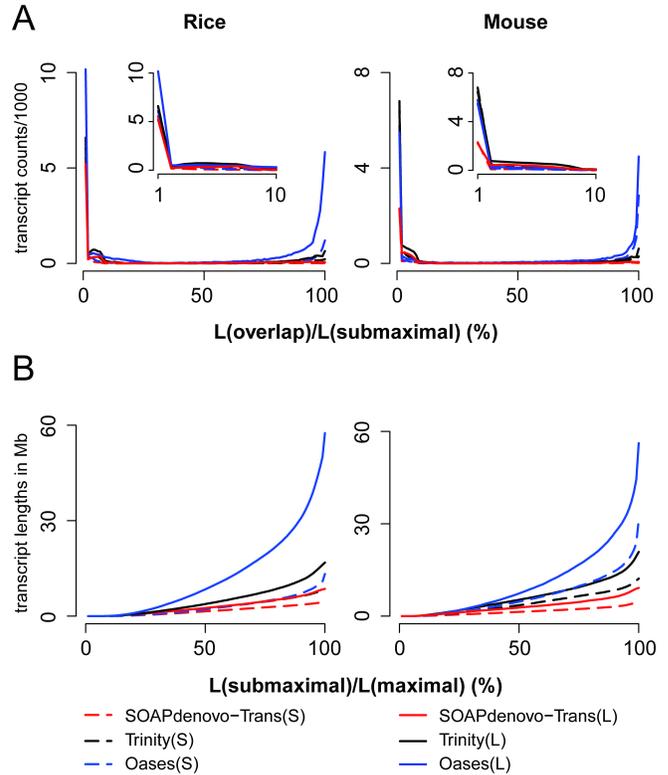

**Fig. 4.** Analysis of alternative splice forms. Given a set of assembled transcripts aligning to the same genome locus, L(submaximal) is the length of any transcript besides the largest while L(maximal) is the length of the largest transcript. L(overlap) is the length of the overlap between the two. Much as with Figure 2 and 3, we show a distribution for the number of transcripts and then a cumulant for the transcript lengths.

To quantify how many genes or isoforms are recovered, one has to choose an arbitrary threshold, say 100% or 95% of the expected length, as shown in Table 3. Note that for this table we only count isoforms recorded in the genome annotations. While it is possible that the transcriptome data contain isoforms that had not been previously discovered, it is equally possible that what we called putative alternative splice forms are really just assembly errors. Counting only what was recorded in the genome annotations indicates, perhaps surprisingly, that Trinity and Oases are not recovering any more isoforms than SOAPdenovo-Trans.

To explore what is causing the assemblers, especially Oases, to generate so many putative alternative splice forms, we show in Fig. 4A a comparison of the submaximal transcripts (*i.e.* all but the largest of the many transcripts that aligned to a particular genome locus) to the maximal transcript. In many cases, there is virtually no overlap between the submaximal and maximal transcripts. What this means is that the assemblers produced non-overlapping fragments of the same isoform. However, in most cases, the overlap to submaximal ratio is one, which means no new exons are recovered, unlike what should have happened for genuine cases of alternative splicing. We have observed that the assemblers often produce multiple artefactual transcripts as a result of minor substitutional errors in the raw input data. All have much the same length, in contrast to the common form of alternative splicing where exons are added or subtracted, resulting in 20 to 10% changes in the transcript lengths (*i.e.* one exon out of 5 to 10 exons in a plant or animal gene). Fig. 4B tests for artefacts of this nature, by plotting the cumulant for the transcript lengths as a function of submaximal-to-maximal lengths. The sharp upticks as the ratios approach one shows that all assemblers do this to varying degrees.

## 4 DISCUSSION

Sequence assembly on real world datasets has always required a many minor algorithmic developments to produce the best results. There is no one magic bullet that solves all of the problems. In this spirit, SOAPdenovo-Trans combined the novel insights from Trinity and Oases, merged them with ideas originally developed for the genome version of SOAPdenovo2, and then added new insights of our own, to produce an algorithm that is demonstrably superior to all previous efforts. This however is unlikely to be the last word in transcriptome assembly. We also tested one of the reference-based assemblers, Cufflinks, and got even better results than SOAPdenovo-Trans. It is unclear to us how much of this improvement can be replicated without recourse to a reference genome, but the results do suggest that there is information in these datasets that, perhaps, with the right algorithm can be recovered.

For example, a multiple *k*-mers strategy may improve transcriptome assembly. Current multiple *k*-mers assembly strategies generally fall into one of two categories: (a) After using different values for *k*-mer assembly, separately, merge the resultant assemblies into one final set. This may construct a more complete transcript set but this may also introduce redundancy. (b) Iterate different *k*-mer *de Bruijn* graph assemblies during contig construction. This strategy potentially makes the best use of reads and paired-end information. Whether or not it is worth the effort to develop such an algorithm depends in part on continuing progress in sequencing technology, since if the promised improvements in read lengths materialize, the nature of the problem will change radically.

## ACKNOWLEDGEMENTS

*Funding*: Much of this work was done in support of National Gene Bank Project of China; Shenzhen Key Laboratory of Transomics Biotechnologies (CXB201108250096A); and the 1000 Plants (1KP) initiative, which is led by GKSW and funded by the Alberta Ministry of Enterprise and Advanced Education, Alberta Innovates Technology Futures (AITF) Innovates Centre of Research Excellence (iCORE), and Musea Ventures.